\documentclass[pra,aps,twocolumn,floatfix,showpacs,tightenlines,superscriptaddress,amsmath,amssymb]{revtex4}
\usepackage{amssymb,amsbsy,epsfig,color,graphicx}

\newcommand{\ket}[1]{|#1\rangle}

\begin{document}

\title{Continuous variable quantum teleportation with non-Gaussian resources}

\author{F. Dell'Anno}
\affiliation{Dipartimento di Matematica e Informatica, Universit\`a
degli Studi di Salerno, Via Ponte don Melillo, I-84084 Fisciano (SA), Italy}
\affiliation{Dipartimento di Fisica, Universit\`a
degli Studi di Salerno, Via S. Allende, I-84081 Baronissi (SA), Italy}
\affiliation{CNR-INFM Coherentia, Napoli, Italy, and INFN Sezione di Napoli,
Gruppo collegato di Salerno, Baronissi (SA), Italy}

\author{S. De Siena}
\affiliation{Dipartimento di Fisica, Universit\`a
degli Studi di Salerno, Via S. Allende, I-84081 Baronissi (SA), Italy}
\affiliation{CNR-INFM Coherentia, Napoli, Italy, and INFN Sezione di Napoli,
Gruppo collegato di Salerno, Baronissi (SA), Italy}

\author{L. Albano Farias}
\affiliation{Dipartimento di Matematica e Informatica, Universit\`a
degli Studi di Salerno, Via Ponte don Melillo, I-84084 Fisciano (SA), Italy}
\affiliation{CNR-INFM Coherentia, Napoli, Italy, and INFN Sezione di Napoli,
Gruppo collegato di Salerno, Baronissi (SA), Italy}

\author{F. Illuminati}
\thanks{Corresponding author. Electronic address: illuminati@sa.infn.it}
\affiliation{Dipartimento di Matematica e Informatica, Universit\`a
degli Studi di Salerno, Via Ponte don Melillo, I-84084 Fisciano (SA), Italy}
\affiliation{CNR-INFM Coherentia, Napoli, Italy, and INFN Sezione di Napoli,
Gruppo collegato di Salerno, Baronissi (SA), Italy}
\affiliation{ISI Foundation for Scientific Interchange,
Viale Settimio Severo 65, 00173 Torino, Italy}

\date{June 25, 2007}

\begin{abstract}
We investigate continuous variable quantum teleportation using
non-Gaussian states of the radiation field as entangled
resources. We compare the performance of different classes
of degaussified resources, including two-mode photon-added  
and two-mode photon-subtracted squeezed states.
We then introduce a class of two-mode squeezed Bell-like states
with one-parameter dependence for optimization. These states interpolate
between and include as subcases different classes of degaussified resources. 
We show that optimized squeezed Bell-like resources
yield a remarkable improvement in the fidelity 
of teleportation both for coherent and nonclassical input states.
The investigation reveals that the optimal non-Gaussian resources
for continuous variable teleportation are those that most closely 
realize the simultaneous maximization of the content of entanglement,
the degree of affinity with the two-mode squeezed vacuum and the, suitably 
measured, amount of non-Gaussianity.
\end{abstract}


\maketitle

\section{Introduction}

Recent theoretical and experimental effort in quantum
optics and quantum information has been focused on the engineering
of highly nonclassical, non-Gaussian states of the radiation field
\cite{PhysRep}, in order to achieve either enhanced properties
of entanglement or other desirable nonclassical features
\cite{PhysRep,KimBS,KitagawaPhotsub,DodonovDisplnumb,VanLoock}.
It has been shown that, at fixed covariance matrix, some of these
properties, including entanglement and distillable secret key rate,
are minimized by Gaussian states \cite{ExtremalGaussian}. In the last
two decades, increasingly sophisticated degaussification protocols
have been proposed, based on photon addition or subtraction
\cite{AgarTara,DeGauss1,DeGauss2,DeGauss3,DeGauss4},
and some of them have been recently experimentally implemented
to engineer non-Gaussian photon-added and photon-subtracted states
starting from Gaussian coherent or squeezed inputs
\cite{ZavattaScience,ExpdeGauss1,ExpdeGauss2}.

Progresses in the theoretical characterization
and the experimental production of non-Gaussian states
are being paralleled by the increasing attention on the
role and uses of non-Gaussian entangled resources in quantum 
information and quantum computation with continuous-variable 
systems \cite{LloydBraunstein}. In particular,
concerning quantum teleportation with continuous variables
\cite{BraunsteinKimble,Furusawa1,VanLoockBraunstein}, it has
been demonstrated that the fidelity of teleportation can
be improved by exploiting suitable deguassifications of
Gaussian resources \cite{Opatrny,Cochrane,Olivares,KitagawaPhotsub}.
Moreover, non-Gaussian cloning of coherent states has been
shown to be optimal with respect to the single-clone fidelity \cite{Cerf}.
Determining the performance of non-Gaussian 
entangled resources may prove useful in a number of concrete 
applications ranging from hybrid quantum computation \cite{Hybrid}
to cat-state logic \cite{Ralph} and, generically, in all quantum computation
schemes based on communication that integrate together qubit degrees of 
freedom for computation with quantum continuous variables for communication 
and interaction \cite{Communication}.

In the present work, we investigate systematically the performance of different
classes of entangled two-mode non-Gaussian states used as resources
for continuous-variable quantum teleportation. In our approach, the entangled
resources are taken to be non-Gaussian {\em ab initio}, and their
properties are characterized by the interplay between continuous-variable (CV) 
squeezing and discrete, single-photon pumping.  
Our first aim is to determine the actual properties of non-Gaussian resources that are 
needed to assure improved performance compared to the Gaussian case. At the same
time, we carry out a comparative analysis between the different non-Gaussian
cases in order to single out those properties that are most relevant to successful 
teleportation. Finally, we wish to understand the role of adjustable free parameters, 
in addition to squeezing, in order to ``sculpture'' and achieve optimized performances 
within the set of non-Gaussian resources. We will show that maximal non-Gaussian improvement 
of teleportation success depends on the nontrivial relations between enhanced entanglement,
suitably measured level of {\em non-Gaussianity}, and the presence of a proper Gaussian
squeezed-vacuum contribution in the non-Gaussian resources at large values of the
squeezing (squeezed-vacuum affinity). We limit the discussion to general issues of 
principle, considering the ideal situation of pure-state resources in the absence of noise
and imperfections. Detailed analysis of realistic situations with mixed-state resources
in the presence of various sources of noise will be discussed elsewhere.

The paper is organized as follows. In Section II we introduce and
describe relevant instances of two-mode entangled non-Gaussian resources, including
squeezed number states and typical degaussified states currently considered in the
literature, such as photon-added squeezed and photon-subtracted squeezed states. 
We show that all of the former, as well 
as the Gaussian two-mode vacuum and squeezed vacuum (twin-beam), can be seen as 
particular subcases of a properly defined class of {\em squeezed Bell-like states}
depending on a continuous angular parameter. 
In Section III, exploiting the unifying formalism of the characteristic function,
we compare the relative performances of non-Gaussian and Gaussian resources 
in the Braunstein-Kimble CV teleportation protocol for different (single-mode) input states, 
Gaussian and non-Gaussian, including coherent and squeezed states, number states, photon-added 
coherent states, and squeezed number states. 
In Section IV we consider the optimization of non-Gaussian performance 
in CV teleportation with respect to the extra angular parameter of squeezed Bell-like states.
We show that maximal teleportation fidelity is achieved using a form of squeezed Bell-like resource 
that differs both from squeezed number and photon-added/subtracted squeezed states. In Section V we
identify the properties that determine the maximization of the teleportation fidelity
using non-Gaussian resources. We find that optimized non-Gaussian resources are those that
come nearest to the simultaneous maximization of three distinct properties: the content of entanglement, 
the amount of (properly quantified) non-Gaussianity, and the degree of {\em ``vacuum affinity''}, i.e.
the maximum, over all values of the squeezing parameter, of the overlap between a non-Gaussian resource 
and the Gaussian twin-beam. Schemes for the experimental
production of optimized squeezed Bell-like resources are illustrated in Section VI. Finally,
in Section VII we present our conclusions and discuss some outlooks about the extension to other 
types of resources, optimized protocols, and applications to realistic situations in the presence 
of noise. 

\section{Non-Gaussian Resources: Characterization and Entanglement Properties}

\label{SecNonGaussSt}

We begin our study by considering some different instances of two-mode entangled non-Gaussian states
obtained by squeezing operations and mechanisms of photon addition/subtraction. 
Let us first introduce the following three classes of (normalized) pure states:
\begin{eqnarray}
|\zeta\,; m_{1}\,, m_{2}\rangle \, &=& \, S_{12}(\zeta) \; |m_{1}\,,m_{2}
\rangle_{12} \,, \label{SqNum}
\\ && \nonumber
\\
|m_{1}^{(+)}\,, m_{2}^{(+)} \,; \zeta \rangle \, &=& \,
\mathcal{N}_{12}^{(+)} a_{1}^{\dag m_{1}} a_{2}^{\dag m_{2}}
\,S_{12}(\zeta) \; |0\,,0 \rangle_{12} \,, \label{PhAddSq}
\\ && \nonumber
\\
|m_{1}^{(-)}\,, m_{2}^{(-)} \,; \zeta \rangle \, &=& \,
\mathcal{N}_{12}^{(-)} a_{1}^{m_{1}} a_{2}^{m_{2}} \, S_{12}(\zeta) \;
|0\,,0 \rangle_{12} \; , 
\label{PhSubSq}
\end{eqnarray}
where $S_{12}(\zeta) = e^{ -\zeta a_{1}^{\dag}a_{2}^{\dag} + \zeta
a_{1}a_{2}}$ is the two-mode squeezing operator, $\zeta=r e^{i\phi}$,
$\mathcal{N}_{12}^{(\cdot)}$ are the normalizations,
and $|m_{1}\,,m_{2} \rangle_{12} \equiv |m_{1}\rangle_{1} \otimes
|m_{2}\rangle_{2}$ is a two-mode Fock state. Eqs. (\ref{SqNum}),
(\ref{PhAddSq}), (\ref{PhSubSq}) define the squeezed number
states, the photon-added squeezed states, and the photon
subtracted squeezed states, respectively. 
Letting $m_{i}^{(\cdot)}=0$, all states reduce to the Gaussian two-mode 
squeezed vacuum, i.e. the {\em twin-beam} $|\zeta \,\rangle \,=\,
S_{12}(\zeta) \, |0,0\rangle_{12}$. The normalization factors can
be easily computed. For instance, if we take the explicit case $m_{i}^{(\cdot)}=1$, 
we have:
\begin{eqnarray}
|\zeta\,; 1\,, 1\rangle \,&=&\, S_{12}(\zeta) \; |1\,,1 \rangle_{12} \,,
\label{SN11}
\\ && \nonumber \\
|1^{(+)} , 1^{(+)} ; \zeta \rangle & = &  \mathcal{N} 
e^{-i\phi} S_{12}(\zeta) \nonumber \\
&& \{ -\tanh r |0 , 0 \rangle_{12} 
+ e^{i\phi} |1 , 1 \rangle_{12} \} \; , 
\label{PAS11} \\ 
&& \nonumber \\
|1^{(-)} , 1^{(-)} ; \zeta \rangle  & = & \mathcal{N} e^{i\phi} 
S_{12}(\zeta) \nonumber \\
&& \{ -|0 , 0 \rangle_{12} +
e^{i\phi} \tanh r |1 , 1 \rangle_{12} \} \; , 
\label{PSS11}
\end{eqnarray}
where $\mathcal{N} = [1 + \tanh^{2} r]^{-1/2}$ is the normalization,
and Eqs. (\ref{PAS11}) and (\ref{PSS11}) have been obtained by
exploiting the two-mode Bogoliubov transformations 
$S_{12}^{\dag}(\zeta)\, a_{i} \, S_{12}(\zeta)=\cosh r \, a_{i}
-e^{i\phi}\sinh r \, a_{j}^{\dag}, \;\, (i\neq j=1,2)$. We remark
that both the photon-added and the photon-subtracted
squeezed states are superpositions of the twin-beam and of the
squeezed number state. However, Eqs. (\ref{PAS11}) and
(\ref{PSS11}) substantially differ for an exchange of the
hyperbolic coefficients: In the limit of vanishing squeezing,
the photon-added squeezed state reduces to a two-mode Fock
state, remaining non-Gaussian, while the photon-subtracted squeezed 
state becomes Gaussian, as it reduces to the two-mode vacuum. 
In fact, all these states are particular instances of what we could
name {\em squeezed Bell-like state}:
\begin{equation}
|\Psi \rangle_{SB} \, = \, S_{12}(\zeta)
\, \{ \cos{\delta} |0\,,0\rangle_{12} \, + \, e^{i \theta} \sin{\delta}
|1\,,1\rangle_{12} \} \; . 
\label{SBL}
\end{equation}
For instance, the squeezed number state (\ref{SN11}) is recovered 
for $\delta = \pi/2$.
 
\begin{figure}[h]
\centering
\includegraphics*[width=7cm]{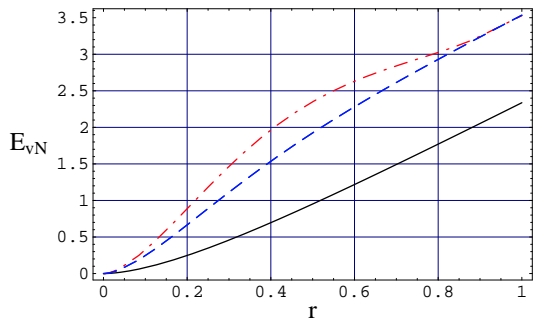}
\caption{(color online) Behavior of the von Neumann entropy $E_{vN}$ for the pure states
Eqs. (\ref{SN11}), (\ref{PAS11}), and (\ref{PSS11}), as a function of 
the modulus $r$ of the squeezing parameter $\zeta$. The upper curve (dot-dashed
line) corresponds to the squeezed number state $|\zeta\,; 1\,,
1\rangle$; the intermediate curve (dashed line) corresponds
equivalently to the photon-added squeezed state $|1^{(+)}\,,
1^{(+)} \,; \zeta \rangle$ and to the photon-subtracted squeezed state
$|1^{(-)}\,, 1^{(-)} \,; \zeta \rangle$. The lower curve corresponds
to the twin-beam $|\zeta\rangle$.}
\label{vonNeumann}
\end{figure}

To quantify the bipartite entanglement of states
(\ref{SN11}), (\ref{PAS11}), (\ref{PSS11}), and (\ref{SBL})
one needs the von Neumann entropy (entropy of entanglement)
$E_{vN}$. For the first three states, this quantity depends
only on the modulus $r$ of the squeezing parameter $\zeta$.
It is plotted in Fig. \ref{vonNeumann} and compared to that
of the twin-beam.
At a given squeezing, all the non-Gaussian states show an entanglement
larger than that of the Gaussian squeezed vacuum. In particular, in the range
of experimentally realistic values $0<r<1$ of the squeezing, the squeezed
number state is the most entangled state. Moreover, the photon-added
and the photon-subtracted squeezed states exhibit
the same amount of entanglement at any $r$.

\begin{figure}[h]
\centering
\includegraphics*[width=7cm]{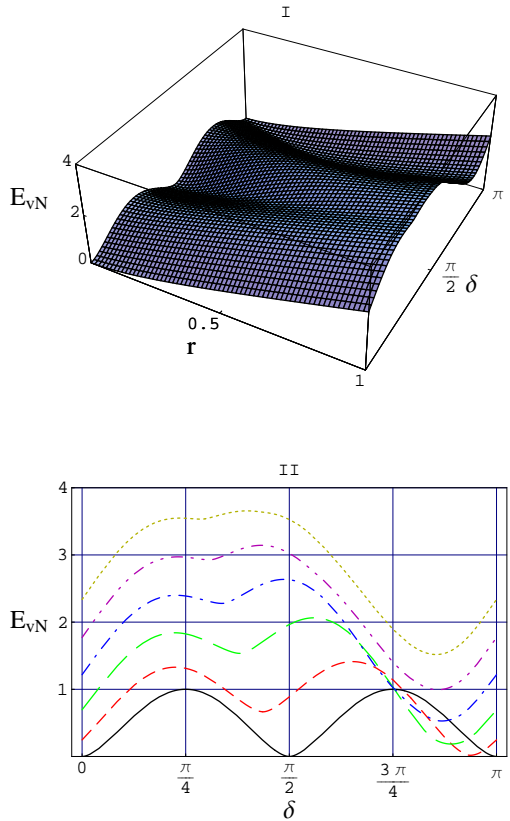}
\caption{(color online) von Neumann entropy $E_{vN}$ for the squeezed Bell-like state
(\ref{SBL}), as a function of $r$ and $\delta$.
Panel I displays the three-dimensional plot of $E_{vN}$. Panel II
displays two-dimensional projections at fixed squeezing strength $r$.
Curves from bottom to top correspond to the different sections of $E_{vN}$ as 
functions of $\delta$ for $r=0,\,0.2,\,0.4,\,0.6,\,0.8,\,1$.}
\label{vonNeumSBell}
\end{figure}

The von Neumann entropy of the squeezed Bell-like state (\ref{SBL})
is plotted in Fig. \ref{vonNeumSBell}. In panel I we plot 
$E_{vN}$ as a function of $r$ and $\delta$. In panel II, we can observe 
how the regular, oscillating behavior of the entropy for
the Bell-like state $(r=0)$ becomes gradually deformed by the optical
pumping $(r\neq 0)$, leading to a peculiar pattern of correlation properties
for the squeezed Bell-like state (\ref{SBL}).

\section{Teleportation in the characteristic function representation}

\label{SecCVTelep}

Quantum teleportation was first proposed by Bennett {\it et al.}
in the discrete variable regime \cite{Bennett}, and later 
experimentally demonstrated in that setting 
\cite{Bouwmeester,DeMartini}. The idea of continuous-variable (CV) teleportation
was put forward by Vaidman \cite{Vaidman}. Some time later
the actual quantum-optical protocol for the teleportation of quadrature
amplitudes of a light field was introduced by Braunstein and Kimble in the 
formalism of the Wigner function \cite{BraunsteinKimble}, and realized by 
Furusawa {\it et al.} soon afterward \cite{Furusawa1,Bowen,Furusawa2}. 
In the standard CV protocol two users, Alice and
Bob, share an entangled state (resource) of modes $A$ and $B$; a
single-mode input state $\ket{in}$, in Alice's possession, is the state
to be teleported. The protocol works as follows: the input mode $in$
and mode $A$ of the entangled resource are
mixed at a $50/50$ beam splitter, yielding the output modes $in'$
and $A'$. A destructive measurement (homodyne) is performed by
Alice on the output modes $in'$ and $A'$. The obtained result is
(classically) communicated to Bob; subsequently, Bob performs a
unitary operation (displacement) on mode $B$, leading to the
teleported state. For a comprehensive review on continuous-variable
quantum teleportation and quantum information processing, see Ref. 
\cite{VanLoockBraunstein}. Various alternative descriptions of the 
original Braunstein-Kimble protocol have been introduced in the 
literature. Among them, we should mention those involving Fock 
state expansion \cite{TelepFormal1}, the coherent state expansion
\cite{TelepFormal2}, and the transfer operator approach
\cite{TelepFormal3}. 

Recently, the CV teleportation protocol has
been described in terms of the characteristic functions of the
quantum states involved (input, resource, and teleported states)
\cite{MarianCVTelep}. This formalism is particularly suited when
dealing with non Gaussian states and resources, because it greatly
simplifies the calculational strategies. Let us denote by
$\rho_{in}$ and $\chi_{in}(\alpha_{in})$, respectively, the
single-mode input state to be teleported and the associated
characteristic function, and by $\rho_{12}$ and $\chi_{12}(\alpha_{1}\,,\alpha_{2})$,
respectively, the entangled two-mode resource, shared by the sender and the receiver,
and its characteristic function. By exploiting the Weyl
expansion, it can be shown that the characteristic function
$\chi_{out}(\alpha_{2})$ of the teleported state has the
factorized form \cite{MarianCVTelep}:
\begin{equation}
\chi_{out}(\alpha_{2}) \,=\, \chi_{in}(\alpha_{2}) \;
\chi_{12}(\alpha_{2}^{*}\,,\alpha_{2}) \; .
\label{telepchiout}
\end{equation}
We should remark the great simplicity, beauty, and power of
this expression, particularly well suited in the study of
teleportation-related subjects.
In order to measure the success probability of a teleportation
protocol, it is convenient to use the fidelity of teleportation 
$\mathcal{F}$. This is a state-dependent quantity that measures
the overlap between the input state $\rho_{in}$ and the output 
(teleported) state $\rho_{out}$, i.e. $\mathcal{F} \, = \, 
Tr[\rho_{in}\rho_{out}]$. In the characteristic-function formalism, 
the fidelity reads
\begin{equation}
\mathcal{F} \,=\,  \frac{1}{\pi} \int d^{2}\mathbf{\lambda} \;
\chi_{in}(\mathbf{\lambda}) \chi_{out}(-\mathbf{\lambda}) \,.
\label{Fidelitychi}
\end{equation}
In the following we will adopt Eq. (\ref{Fidelitychi}) to analyze
the efficiency of the CV teleportation protocol for different classes
of input states and non Gaussian entangled resources.

Let us first compute the symmetrically ordered characteristic function
for the squeezed-number states, the photon-added, and the photon-subtracted
squeezed states Eqs. (\ref{SN11}), (\ref{PAS11}), (\ref{PSS11}). Being two-mode
states, their characteristic function is of the form  
$\chi (\alpha_{1}\,,\alpha_{2}) \, = \,
Tr[D_{1}(\alpha_{1})\, D_{2}(\alpha_{2})\, \rho]$, where
$D_{i}(\alpha_{i})$ is the displacement operator corresponding to
mode $i$, and $\rho$ is the density operator associated to the state. 
We will make use of the relation
\begin{equation}
\langle m| D(\alpha) |n \rangle \,=\,
\left(\frac{n!}{m!}\right)^{1/2}\alpha^{m-n}e^{-\frac{1}{2}|\alpha|^{2}}
L_{n}^{(m-n)}(|\alpha|^{2}) \,,
\end{equation}
where $L_{n}^{(m-n)}(\cdot)$ is the associate Laguerre polynomial.
The characteristic function for the state $|\zeta\,; 1\,, 1\rangle$ is
\begin{equation}
\chi_{SN}^{(1,1)}(\alpha_{1} , \alpha_{2}) =
\chi_{S}(1-|\xi_{1}|^{2})(1-|\xi_{2}|^{2}) \; , 
\label{chiSN}
\end{equation}
where 
\begin{equation}
\chi_{S}(\alpha_{1}\,,\alpha_{2}) \, = \,
e^{-\frac{1}{2}(|\xi_{1}|^{2} + |\xi_{2}|^{2})} 
\label{chiSS}
\end{equation}
is the characteristic function of the two-mode squeezed state,
the standard reference Gaussian resource, and the implicit dependence 
on $\alpha_{i}$ stems from the relations
\begin{equation}
\xi_{k} = \alpha_{k} \cosh r
+\alpha_{l}^{*} e^{i\phi} \sinh r , \; (k,l=1,2 ; \, k \neq l). 
\label{defxir}
\end{equation}
The characteristic functions for the states $|1^{(+)}\,, 1^{(+)}
\,; \zeta \rangle$ and $|1^{(-)}\,, 1^{(-)} \,; \zeta \rangle$ read,
respectively,
\begin{eqnarray}
\chi_{PAS}^{(1,1)}(\alpha_{1} , \alpha_{2}) & = &
{\mathcal{N}}^{2}\chi_{S}
\{ \tanh^{2} r - 2 \tanh r Re[e^{-i \phi} \xi_{1} \xi_{2}] \nonumber \\
&& \nonumber \\
& + & (1-|\xi_{1}|^{2}) (1-|\xi_{2}|^{2}) \} \; ,
\label{chiPAS}
\\ && \nonumber \\
\chi_{PSS}^{(1,1)}(\alpha_{1} , \alpha_{2}) & = & 
{\mathcal{N}}^{2}\chi_{S} \{ 1 - 2 \tanh r Re[e^{-i \phi} 
\xi_{1} \xi_{2}] \nonumber \\ 
&& \nonumber \\
& + & \tanh^{2} r (1-|\xi_{1}|^{2}) (1-|\xi_{2}|^{2}) \} \; .
\label{chiPSS}
\end{eqnarray}
Comparing Eq. (\ref{chiSS}) with Eqs. (\ref{chiSN}), (\ref{chiPAS}),
and (\ref{chiPSS}), we see that the polynomial non-Gaussian forms are 
always modulated by a Gaussian factor that coincides exactly with the
squeezed-state characteristic function Eq. (\ref{chiSS}).

\section{Teleportation with non-Gaussian resources}

In this Section we will compare the behavior of the fidelity for
different input states by making use of the non-Gaussian entangled
resources (\ref{SN11}), (\ref{PAS11}), and
(\ref{PSS11}). The analysis will be carried out in terms of the
entangling parameter $\zeta$ common to all resources. 
The following single-mode input states will be
considered: coherent states $|\beta\rangle$; squeezed states
$|\varepsilon\rangle \,=\, S(\varepsilon)|0\rangle$, with
$S(\varepsilon) \,=\, \exp\left\{-\frac{1}{2}\varepsilon a^{\dag
2}+\frac{1}{2}\varepsilon^{*} a^{2}\right\}$ $(\varepsilon =
e^{i \varphi}s)$; single-photon Fock states $|1\rangle$; 
photon-added coherent states $(1+|\beta|^{2})^{-1/2}a^{\dag}|\beta\rangle$;
and squeezed single-photon Fock states $S(\varepsilon)|1\rangle$. The
teleportation implemented with the two-mode squeezed Gaussian resource 
$|\zeta\rangle = S_{12}(\zeta) \, |0,0\rangle_{12}$ as
entangled resource will be used as standard reference for comparison. Let us
first consider the behavior of the fidelity for the Gaussian input
states $|\beta\rangle$ and $|\varepsilon\rangle$, whose
characteristic functions read
\begin{eqnarray}
\chi_{coh}(\alpha_{in}) \,&=&\,
e^{-\frac{1}{2}|\alpha_{in}|^{2}+2i Im[\alpha_{in}\beta^{*}]}
\; , 
\label{chicohin}
\\ 
&& \nonumber \\
\chi_{sq}(\alpha_{in}) \,&=&\, e^{-\frac{1}{2}|\xi_{in}|^{2}} \; ,
\label{chisqin}
\end{eqnarray}
where
\begin{equation}
\xi_{in} \,=\, \alpha_{in} \cosh s \,+\,
\alpha_{in}^{*} e^{i \varphi}\sinh s \; . 
\label{cazzin}
\end{equation}
Let us remark that the fidelity is analytically computable for the
the classes of input states and entangled resources considered, as 
the integral in Eq. (\ref{Fidelitychi}) can be exactly calculated in terms of
finite sums of Gaussian averages. 

\begin{figure}[h]
\centering
\includegraphics*[width=7cm]{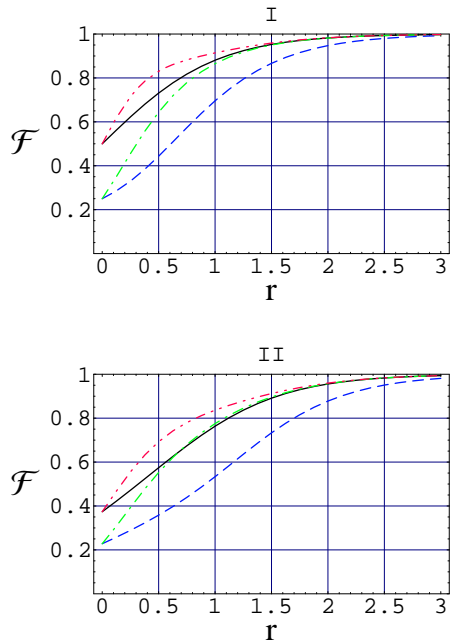}
\caption{(color online) Fidelity of teleportation
$\mathcal{F}$, as a function of the squeezing parameter 
$r$, with $\phi=\pi$, for input coherent states $|\beta\rangle$
(Panel I) and input squeezed states $|\varepsilon\rangle$ (Panel II).
Comparison is given for different two-mode entangled resources: $(a)$ squeezed 
state (full line); $(b)$ squeezed number state (dashed line);
$(c)$ photon-added squeezed state (dot-dashed line); $(d)$ photon-subtracted 
squeezed state (double-dotted, dashed line). In plot I the value of $\beta$ is arbitrary.
In plot II the squeezing parameter $\varepsilon$ of the input state is fixed at modulus
$s=0.8$ and phase $\varphi=0$.}
\label{FigFidGaussIn}
\end{figure}

In Fig. \ref{FigFidGaussIn} we plot the
fidelity $\mathcal{F}$ for input coherent states $|\beta\rangle$ (Panel I), 
and input squeezed states (Panel II).
We see that in both cases, the choice of the photon-subtracted squeezed state
(\ref{PSS11}) as entangled resource is the most convenient one. It
corresponds to the highest value of the fidelity $\mathcal{F}$ for any
fixed value of the squeezing $r$ (or, equivalently, of the energy) in the
realistic range $[0,1]$. On the contrary, the choice of the squeezed number state (\ref{SN11})
as entangled resource is the least convenient, yielding the poorest performance
even when compared to the Gaussian squeezed resource. Finally, regarding the 
use of the photon-added squeezed state (\ref{PAS11}) as entangled resource, 
it allows for a very modest improvement in the fidelity compared to the Gaussian
resource, and this only for a small interval of values around $r=1$.

\begin{figure}[h]
\centering
\includegraphics*[width=7cm]{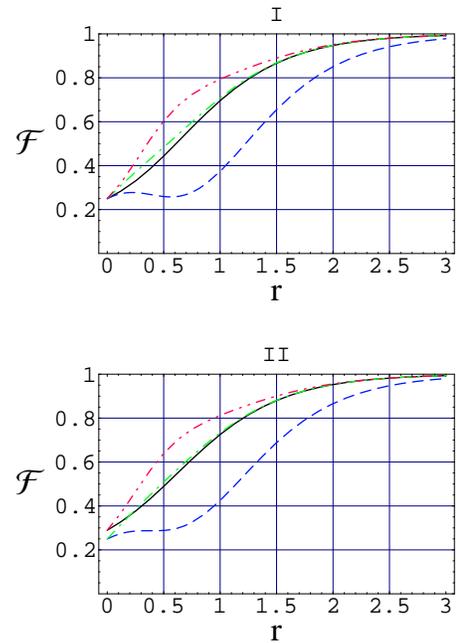}
\caption{(color online) Behavior of the fidelity of teleportation $\mathcal{F}$
as a function of the squeezing parameter $r$, with $\phi=\pi$, for two different
non-Gaussian input states: The Fock state $|1\rangle$ (Panel I), and the
photon-added coherent state $(1+|\beta|^{2})^{-1/2}a^{\dag}|\beta\rangle$ (Panel II). 
We compare the performances obtained by using different two-mode
entangled Gaussian and non-Gaussian resources:
$(a)$ squeezed state (full line);
$(b)$ squeezed number state (dashed line);
$(c)$ photon-added squeezed state (dot-dashed line);
$(d)$ photon-subtracted squeezed state (double-dotted, dashed line).
In Panel II the value of the coherent amplitude of the input photon-added
coherent state is fixed at $\beta=0.3$.}
\label{FigFidNonGaussIn1}
\end{figure}

Let us now consider the case of non-Gaussian input states
$|1\rangle$, $(1+|\beta|^{2})^{-1/2}a^{\dag}|\beta\rangle$, and
$S(\varepsilon)|1\rangle$, whose characteristic functions read,
respectively:
\begin{eqnarray}
\chi_{F}(\alpha_{in}) & = & e^{-\frac{1}{2}|\alpha_{in}|^{2}} \,
(1-|\alpha_{in}|^{2}) \; , 
\label{chiFockin}
\\ && \nonumber \\
\chi_{pac}(\alpha_{in}) & = & (1+|\beta|^{2})^{-1}
e^{-\frac{1}{2}|\alpha_{in}|^{2}+3i Im[\alpha_{in}\beta^{*}]}(1+
\nonumber \\ 
&& \nonumber \\
&& |\beta|^{2}-|\alpha_{in}|^{2}+2i
Im[\alpha_{in}\beta^{*}]) \, , 
\label{chiphaddcohin} \\ 
&& \nonumber \\
\chi_{sqF}(\alpha_{in}) & = &
e^{-\frac{1}{2}|\xi_{in}|^{2}}(1-|\xi_{in}|^{2}) \; .
\label{chisqFockin}
\end{eqnarray}
In Fig. \ref{FigFidNonGaussIn1} we plot the fidelity of teleportation 
for two non-Gaussian input states: the single-photon Fock state 
Eq. (\ref{chiFockin}) (Panel I), and the photon added coherent state 
Eq. (\ref{chiphaddcohin}) (Panel II).
In Panel I, we observe that both the photon-added and photon-subtracted
two-mode squeezed resources (\ref{PAS11}) and
(\ref{PSS11}) lead to an improvement of the fidelity with respect
to the squeezed Gaussian resource. The photon-subtracted squeezed 
state again performs better than the photon-added one, and the 
squeezed number state yields the poorest performance when compared
to the other Gaussian and non-Gaussian resources. From Panel II 
we see that once more the photon-subtracted resource yields the best
performance at any fixed squeezing, and that the photon-added squeezed
state allows for a very modest improvement in the fidelity 
with respect to the squeezed Gaussian reference, above a threshold
value of the squeezing parameter. 

\begin{figure}[h]
\centering
\includegraphics*[width=7cm]{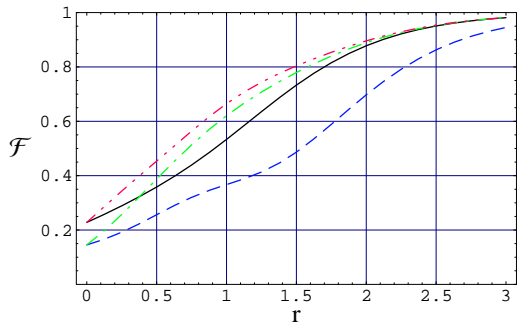}
\caption{(color online) Behavior of the fidelity of teleportation $\mathcal{F}$
as a function of the squeezing parameter $r$, with $\phi=\pi$, for the 
squeezed Fock input state $S(s)|1\rangle$, using different two-mode
Gaussian and non-Gaussian entangled resources:
$(a)$ squeezed state (full line);
$(b)$ squeezed number state (dashed line);
$(c)$ photon-added squeezed state (dot-dashed line);
$(d)$ photon-subtracted squeezed state (double-dotted, dashed line).
The value of $s$ is fixed to $s=0.8$.}
\label{FigFidNonGaussIn2}
\end{figure}

In Fig. \ref{FigFidNonGaussIn2}
we compare the fidelity of teleportation $\mathcal{F}$ 
for the case of a squeezed Fock input state and different Gaussian
and non-Gaussian entangled resources. Comparing with Panels I and II of
Fig. \ref{FigFidNonGaussIn1}, we see that the qualitative behaviors
are very similar to the two previous examples of non-Gaussian input states.
From all of the above investigations, we find that the photon-subtracted squeezed
state (\ref{PSS11}) is always to be preferred as entangled resource
compared either to the Gaussian ones or to non-Gaussian states that
are obtained by combining squeezing and photon pumping.
The reason explaining this result will become clear in the next Sections when we will
discuss a general class of states that include as particular cases all 
the resources introduced so far, and that allow to single out some properties
that are necessary in order to optimize the success of teleportation. 

Before ending this Section, it is worth remarking that the (non-Gaussian) 
two-mode photon-subtracted squeezed state can formally
be defined as the first-order truncation of the (Gaussian) two-mode squeezed
state. Let us first consider the twin-beam $|-2r\rangle \,=\,
S_{12}(-2r)|0,0\rangle_{12}$. Such a state can be written as
$|-2r\rangle \,=\, S_{12}(-r)S_{12}(-r)|0,0\rangle_{12} \propto
S_{12}(-r) \sum_{n=0}^{\infty}\tanh^{n} r|n,n\rangle_{12}$.
Therefore, truncating the series in the last expression at $n=1$,
one recovers the state (\ref{PSS11}), with $\phi=\pi$, i. e.
$|1^{(-)}\,, 1^{(-)} \,; -r \rangle \, \propto \, S_{12}(-r) \{
|0\,,0 \rangle_{12} + \tanh r |1\,,1 \rangle_{12} \}$. Moreover,
expression  (\ref{PSS11}) coincides with that of the photon-subtracted 
state introduced in Ref. \cite{KitagawaPhotsub} when one reduces to the
ideal case of a beam splitter with unity transmittance.

\section{Teleportation with optimized non-Gaussian resources}

\label{SecOptTelep}

In this Section we seek to optimize the fidelity of teleportation, given 
the Vaidman-Braunstein-Kimble protocol, by introducing a class of entangled 
non-Gaussian resources that include as particular cases non-Gaussian photon-added
and photon-subtracted squeezed state, squeezed number states, and Gaussian two-mode
squeezed states and two-mode vacua. We name these states {\em squeezed Bell-like
states}; their general expression reads:
\begin{equation}
|\psi\rangle_{SB} = [c_{1}^{2}+c_{2}^{2}]^{-1/2} S_{12}(\zeta)
\{c_{1} |0,0 \rangle_{12} + e^{i \theta} c_{2}
|1,1 \rangle_{12}\} , 
\label{squeezsuperpos}
\end{equation}
where the $c_{i}$s are real constants. The crucial qualitative
aspect of superpositions (\ref{squeezsuperpos}) lies in their
intrinsic nonclassicality, even at vanishing squeezing: In the 
limit $r \rightarrow 0 $ and for suitable choices of the parameters
$c_{1}$, $c_{2}$, and $\theta$, state (\ref{squeezsuperpos})
reduces to a proper, maximally entangled, Bell state of two
qubits. On the contrary, in the limit of vanishing squeezing,
the two-mode states (\ref{PAS11}) and (\ref{PSS11}) reduce to two different,
factorized (disentangled) limits, respectively the (non-Gaussian)
first excited Fock state and the (Gaussian) two-mode vacuum.

States (\ref{PAS11}) and (\ref{PSS11}) can always be obtained
as particular cases of state (\ref{squeezsuperpos}). For instance,  
fixing the choice $c_{1}=-1$, $c_{2}=\tanh r$, $\theta=\phi$,
the states (\ref{squeezsuperpos}) and (\ref{PSS11}) coincide.
Moreover,
Eq. (\ref{squeezsuperpos}) can be obtained as a superposition of
Eqs. (\ref{PAS11}) and (\ref{PSS11}). A discussion of schemes for
the experimental generation of states (\ref{squeezsuperpos}) is
reported in Section \ref{ExpGeneration}. The characteristic
function associated to the squeezed Bell-like state
(\ref{squeezsuperpos}) reads:
\begin{eqnarray}
\chi_{SB} & = & [c_{1}^{2} +
c_{2}^{2}]^{-1}e^{-\frac{1}{2}(|\xi_{1}|^{2} + |\xi_{2}|^{2})} 
\{ c_{1}^{2} + 2 c_{1}c_{2} Re[e^{i \theta} \xi_{1} \xi_{2}] \nonumber \\
&& \nonumber \\
& + & c_{2}^{2}(1-|\xi_{1}|^{2}) (1-|\xi_{2}|^{2}) \} \, ,
\label{chiSSN}
\end{eqnarray}
where the independent variables $\xi_{k}$ are defined according
to Eq. (\ref{defxir}). 

\begin{figure}[h]
\centering
\includegraphics*[width=7cm]{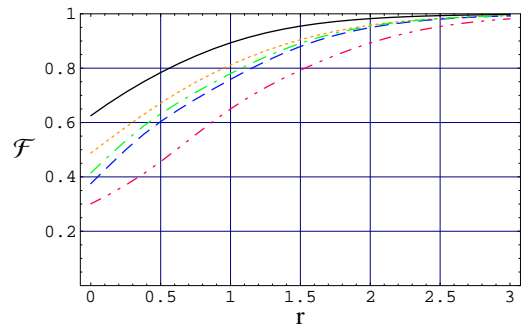}
\caption{(color online) Behavior of the fidelity of teleportation
$\mathcal{F}(r,\phi,\delta, \theta)$ associated to the
squeezed Bell-like resource (\ref{squeezsuperpos}) with
$\phi=\pi$, $\delta=\frac{\pi}{4}$, $\theta=0$, plotted
as a function of the squeezing parameter $r$
for the following input states:
$(a)$ coherent state (full line);
$(b)$ squeezed state $|s\rangle = S(s)|0\rangle$, with $s=0.8$ (dotted line);
$(c)$ Fock state $|1\rangle$ (dashed line);
$(d)$ photon-added coherent state $(1+|\beta|^{2})^{-1/2}a^{\dag}|\beta\rangle$,
with $\beta=0.3$ (dot-dashed line);
$(e)$ squeezed Fock state  $|s\rangle = S(s)|1\rangle$,
with $s=0.8$ (double-dotted, dashed line).}
\label{FigFidsqsuperpos1}
\end{figure}

Exploiting Eqs. (\ref{telepchiout}) and 
(\ref{Fidelitychi}), the expression for the fidelity of teleportation 
$\mathcal{F}$ can be determined analytically for all cases of
of entangled resources of the form (\ref{chiSSN}) and different
input states. In order to
simplify notations, let us introduce the parameterization $c_{1}
= \cos{\delta}$, $c_{2} = \sin{\delta}$. For each given input
state, the
analytic expression for the fidelity will be a function of the
independent parameters $r$, $\phi$, $\delta$, and $\theta$, i.e.
$\mathcal{F}=\mathcal{F}(r,\phi,\delta, \theta)$. For instance,
at finite squeezing and for $\delta=\frac{\pi}{4}$ and $\theta=0$, 
state (\ref{squeezsuperpos}) reduces to a squeezed Bell state and
we may assess analytically the performance of such an entangled
resource as far as teleportation is concerned. In
Fig. \ref{FigFidsqsuperpos1} we show the behavior of 
the fidelity as a function of
the squeezing parameter $r$, with $\phi=\pi$,
$\delta=\frac{\pi}{4}$, $\theta=0$, for the five different
input states considered in the previous Section.
It is straightforward to observe that the squeezed Bell state
(\ref{squeezsuperpos}) with $\delta=\frac{\pi}{4}$ and $\theta=0$,
when used as entangled resource, leads to a relevant improvement 
of the performance, compared to all other Gaussian and non-Gaussian 
resources that we have investigated in the previous Section.
We do not report the explicit analytic expressions of the
fidelities associated to the squeezed Bell-like resource and
to each input state, because they are rather long 
and cumbersome. For the same reason we have not reported
the explicit expressions associated to the other non-Gaussian
entangled resources in the previous Section. In fact, besides 
not being particularly illuminating, reporting the explicit 
expressions is not really needed once the explicit analysis
has established that all fidelities are monotonically increasing
functions of the squeezing parameter $r$ at maximally fixed
phase $\phi=\pi$. Therefore, in the following we will assume 
$\phi=\pi$ and, moreover, $\theta=0$, because one can check 
that nonvanishing values of $\theta$ do not lead to any
improvement of the fidelity. 

\begin{figure}[h]
\centering
\includegraphics*[width=7cm]{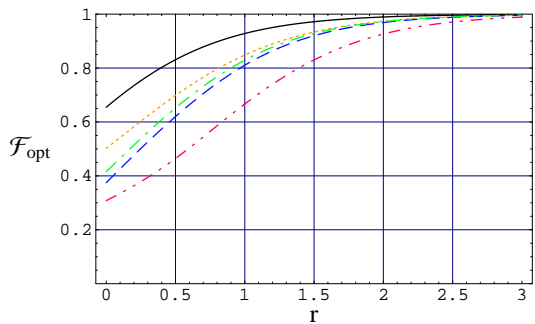}
\caption{(color online) Plot of $\mathcal{F}_{opt}(r)$
as a function of $r$ for the following input states:
$(a)$ coherent state (full line);
$(b)$ squeezed vacuum $|s\rangle = S(s)|0\rangle$, with $s=0.8$ (dotted line);
$(c)$ single-photon Fock state $|1\rangle$ (dashed line);
$(d)$ photon-added coherent state $(1+|\beta|^{2})^{-1/2}a^{\dag}|\beta\rangle$,
with $\beta=0.3$ (dot-dashed line);
$(e)$ squeezed Fock state  $|s\rangle = S(s)|1\rangle$,
with $s=0.8$ (double-dotted, dashed line).}
\label{FigOptimalFid}
\end{figure}

Having established such a framework, we can proceed
to maximize, for each different input state, the fidelity
$\mathcal{F}(r,\pi,\delta, 0)$ over the Bell-superposition
angle $\delta$. At fixed squeezing $r=\tilde{r}$, we define 
the optimized fidelity as
\begin{equation}
\mathcal{F}_{opt}(\tilde{r}) \,=\, \max_{\delta} \;
\mathcal{F}(\tilde{r},\pi,\delta, 0) \; . \label{FidOptim}
\end{equation}
For instance, in the case of input coherent state, the maximization of
$\mathcal{F}(r,\pi,\delta, 0)$, at fixed $r$, leads to the following
determination for the optimal Bell-superposition angle
$\delta_{max}^{(c)}$:
\begin{equation}
\delta_{max}^{(c)} \,=\, \frac{1}{2}\arctan [1+e^{-2r}] \; ,
\label{deltaoptC}
\end{equation}
while for an input single-photon Fock state, one finds:
\begin{equation}
\delta_{max}^{(F)} \,=\, \frac{1}{2}\arctan \Big[
\frac{e^{-2r}(1-e^{2r}+e^{4r}+3e^{6r})}{3(e^{2r}-1)^{2}} \Big] \; .
\label{deltaoptF}
\end{equation}
Finally, in Fig. \ref{FigOptimalFid} we report 
the behavior of the optimized fidelities $\mathcal{F}_{opt}(r)$
as functions of $r$ for all the considered input states.
A relevant improvement of the fidelity is observed in all cases, 
even at vanishing squeezing, due to the persistent nonclassicality
of the squeezed Bell-like entangled resource in the limit
$r \rightarrow 0$. 

\begin{figure}[h]
\centering
\includegraphics*[width=7cm]{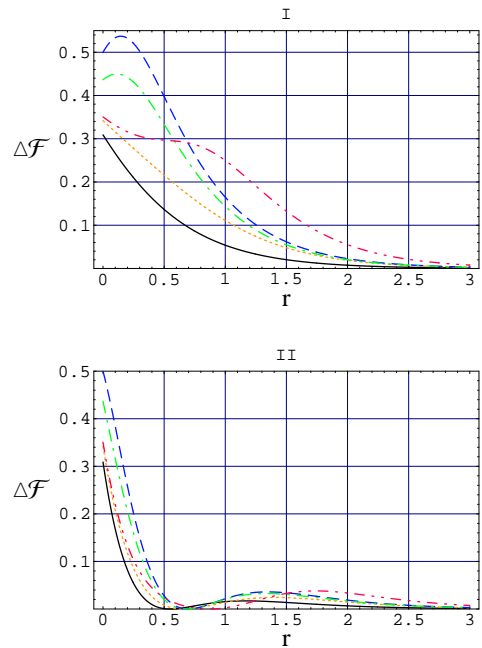}
\caption{(color online) Behavior of the relative fidelity
$\Delta\mathcal{F} \,=\, \mathcal{F}_{opt}(r) - \mathcal{F}_{ref}(r,\pi)$
as a function of $r$, for the following input states:
$(a)$ coherent state (full line);
$(b)$ squeezed state $|s\rangle = S(s)|0\rangle$, with $s=0.8$ (dotted line);
$(c)$ single-photon Fock state $|1\rangle$ (dashed line);
$(d)$ photon-added coherent state $(1+|\beta|^{2})^{-1/2}a^{\dag}|\beta\rangle$,
with $\beta=0.3$ (dot-dashed line);
$(e)$ squeezed Fock state  $|s\rangle = S(s)|1\rangle$,
with $s=0.8$ (double-dotted, dashed line).
In Panel I the reference resource is the twin-beam;
in Panel II the reference resource is the two-mode
photon-subtracted squeezed state.}
\label{FigDeltaFid}
\end{figure}

In order to quantify the increase in the probability of success
for teleportation, we look at the percent increase in fidelity 
relative to a fixed reference. We thus define the difference between 
the optimized fidelity $\mathcal{F}_{opt}(r)$ and the reference
fidelity $\mathcal{F}_{ref}(r,\pi)$, and normalize this difference
by $\mathcal{F}_{ref}(r,\pi)$:
\begin{equation}
\Delta\mathcal{F} (r) \,=\, \frac{\mathcal{F}_{opt}(r) -
\mathcal{F}_{ref}(r,\pi)}{\mathcal{F}_{ref}(r,\pi)} \; ,
\label{deltafidelity}
\end{equation}
where the reference fidelity is fixed to be the one associated to 
a given entangled resource. 
In Fig. \ref{FigDeltaFid}, the relative fidelity $\Delta\mathcal{F}(r)$ 
is plotted as a function of $r$ for two different choices of the reference 
resource. In Panel I, the reference resource is the Gaussian twin-beam; 
in Panel II the reference resource is the non-Gaussian two-mode
photon-subtracted squeezed state.
From Panel I, as expected, we see that, at fixed squeezing (or fixed
energy), the optimized non-Gaussian squeezed Bell-like resource 
leads to a strong percent enhancement of the teleportation 
fidelity (up to more than $50\%$) with respect to that attainable exploiting
the standard Gaussian twin-beam, for every value of $r$. Obviously,
in the asymptotic limit of very large squeezing, the two resources 
converge to perfect teleportation efficiency. 
Panel II shows that use of the optimized squeezed Bell-like entangled resource 
(\ref{squeezsuperpos}) leads to a significant advantage with respect to 
exploiting the photon-subtracted squeezed state resource for low 
values (up to $r \simeq 0.5$) of the squeezing. Moreover, the different
curves corresponding to the different input states, exhibit the same 
qualitative behavior. Starting from large, nonvanishing values,
$\Delta\mathcal{F} (r)$ decreases monotonically and vanishes at different
points in the interval $[0.5 \leq r \leq 0.9]$. It then exhibits revivals 
with different peaks at intermediate values of the squeezing, before 
vanishing asymptotically for large values of $r$. It can be checked that
for values $r=\bar{r}$ such that $\Delta\mathcal{F} (\bar{r}) = 0$, the squeezed
Bell-like state (\ref{squeezsuperpos}) and the photon-subtracted
squeezed state (\ref{PSS11}) coincide.

\section{Understanding optimization: entanglement, non-Gaussianity, and squeezed-vacuum affinity}

\begin{figure}[h]
\centering
\includegraphics*[width=7cm]{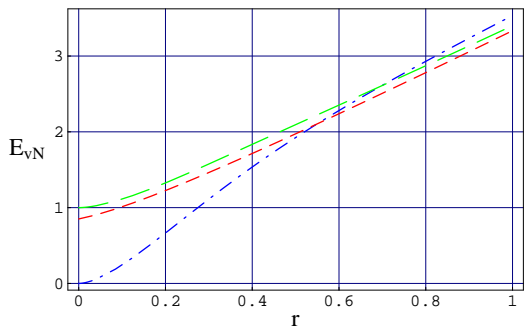}
\caption{(color online) Entropy of entanglement $E_{vN}$ for the squeezed Bell-like state
(\ref{squeezsuperpos}), as a function of $r$, with $\delta$ fixed by Eqs. 
(\ref{deltaoptC}) and (\ref{deltaoptF}). Dashed line: $\delta = \delta_{max}^{(C)}$;
Long dashed line: $\delta = \delta_{max}^{(F)}$. The entropy of the states 
(\ref{PAS11}) and (\ref{PSS11}) is reported as well for comparison (dot-dashed line).}
\label{vonNeumSBelloptCF}
\end{figure}

In this Section we will investigate and determine the properties
that appear to be necessary to achieve maximal teleportation success
with non-Gaussian entangled resources. To this end, we analyze the 
entanglement and the non-Gaussianity of the squeezed Bell-like states
and compare them with those of the photon-added and photon-subtracted
squeezed states. 
In Fig. \ref{vonNeumSBelloptCF} we show the behavior of the von
Neumann entropy $E_{vN}$ for two different squeezed Bell-like resources, 
respectively the one optimized for the teleportation of an input coherent state,
i.e. with $\delta$ given by expression (\ref{deltaoptC}), 
and the one optimized for the teleportation of an input single-photon Fock state, 
i.e. with $\delta$ given by the expression (\ref{deltaoptF}). This behavior is
compared with that of the von Neumann entropy of the photon-added and the photon-subtracted 
squeezed states (the two states have the same degree of entanglement at given
squeezing).
The intersections between the curves
correspond to the values $\bar{r}$ for which the squeezed
Bell-like state reduces to a photon-subtracted or to a photon-added
squeezed state. 
It is then important to observe that in the range $0 < r < \bar{r}$, 
in which the fidelity of teleportation using optimized Bell-like
resources is always maximal (see Fig. \ref{FigDeltaFid}, Panel II), 
the entanglement of the squeezed Bell-like state is always larger
than that of the photon-subtracted (as well as photon-added) 
squeezed states. Therefore, a partial explanation of the better
performance of squeezed Bell-like resources lies in their higher
degree of entanglement compared to other non-Gaussian resources.
However, from the graphs one can see that there are situations in
which the entanglement of photon-added and/or subtracted resources
is larger but, nevertheless, the fidelity of teleportation is still
below the one associated to a squeezed Bell-like resources. Entanglement
is thus not the only characterizing property in order to compare the
performances of different non-Gaussian resources.

\begin{figure}[h]
\centering
\includegraphics*[width=7cm]{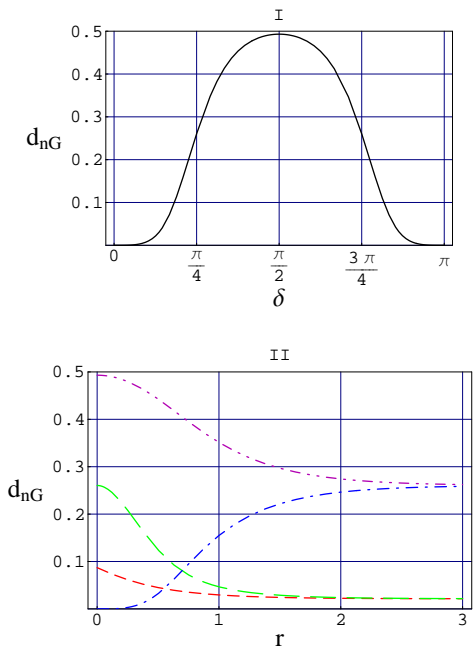}
\caption{(color online) Non-Gaussianity measure $d_{nG}$ for the squeezed Bell-like state
(\ref{squeezsuperpos}).
In panel I we plot $d_{nG}$ for the state (\ref{squeezsuperpos}) as a
function of $\delta$, and for arbitrary $r$.
Panel II reports $d_{nG}$ for the state (\ref{squeezsuperpos}) as a
function of $r$ and for $\delta$ fixed at the optimized values
$\delta = \delta_{max}^{(C)}$ (dashed line), and
$\delta = \delta_{max}^{(F)}$ (long dashed line), see
Eqs. (\ref{deltaoptC}) and (\ref{deltaoptF}).
For comparison the measures for the states (\ref{PAS11}) (double dotted, dashed line)
and (\ref{PSS11}) (dot-dashed line) are also reported.}
\label{nonGaussMeas}
\end{figure}

From the above discussion, it is natural to look at a quantification of 
the non-Gaussian character of different resources, in order to compare 
their performances. Clearly, the 
subtle problem here is to define a reasonable ``measure'' of non-Gaussianity, 
endowed with some nontrivial operative meaning. Recently, inspired by the
analysis of Wolf, Giedke, and Cirac on the extremality of Gaussian states
\cite{ExtremalGaussian} at fixed covariance matrix, a measure of
non-Gaussianity has been introduced in terms of the Hilbert-Schmidt distance
between a given non-Gaussian state and a reference Gaussian state with
the same covariance matrix \cite{GenoniNonGaussy}. Given a generic state
$\rho$, its non-Gaussian character can be quantified through the
distance $d_{nG}$ between $\rho$ and the reference Gaussian state $\rho_{G}$,
defined according to the following relation:
\begin{equation}
d_{nG} = \frac{Tr[(\rho-\rho_{G})^{2}]}{2Tr[\rho^{2}]} =
\frac{Tr[\rho^{2}]+Tr[\rho_{G}^{2}]-2Tr[\rho \rho_{G}]}{2
Tr[\rho^{2}]} ,
\label{nonGaussianity}
\end{equation}
where, as already mentioned, the Gaussian state $\rho_{G}$ is
completely determined by fixing for it the same covariance matrix
and the same first order mean values of the quadrature operators
associated to state $\rho$. Using this definition, in Fig. \ref{nonGaussMeas} 
we report the behavior of the {\it non-Gaussianity} 
$d_{nG}$ for the squeezed Bell-like state (\ref{squeezsuperpos}).
The quantity $d_{nG}$ depends only on the parameter $\delta$ (see
Panel I), as the non-Gaussianity of the state cannot change under 
symplectic squeezing operations. For $\delta$ in the interval
$[0,\pi]$, $d_{nG}$ attains its maximum at $\delta=\frac{\pi}{2}$:
At that point, the Bell-like state reduces to a Fock state. In
fact, as expected, a (squeezed) number state must be more strongly
non-Gaussian than a (squeezed) superposition of the vacuum and of a
Fock state. In Panel II, we report the behavior of $d_{nG}$ for the 
squeezed Bell-like resources optimized for the teleportation 
of a coherent state input and a single-photon Fock state input, i.e.,
respectively with $\delta = \delta_{max}^{(C)}$, and 
$\delta = \delta_{max}^{(F)}$. For comparison, we plot as well the 
non-Gaussianity $d_{nG}$ for the photon-added and the photon-subtracted 
squeezed states. 
The intersection points occur once again at the points
$\bar{r}$ where the squeezed Bell-like states reduce to the 
photon-subtracted squeezed states. For $r$ in
the range $[0,\bar{r}]$, the optimized squeezed Bell-like resources
are not only highly more entangled but as well strongly more non-Gaussian 
than the photon-subtracted squeezed states. One should note that
$\lim_{r\rightarrow +\infty} \delta_{max}^{(C)}
\,=\,\lim_{r\rightarrow +\infty} \delta_{max}^{(F)} \,=\, 1$.
Therefore, for very large squeezing the two optimized squeezed Bell-like resources
tend to the state $S_{12}(-r)\{\cos \frac{\pi}{8}|0,0\rangle_{12}
+ \sin \frac{\pi}{8}|1,1\rangle_{12}\}$, which exhibits a dominating
Gaussian component. On the other hand, for
large $r$, the squeezed photon-added and photon-subtracted
squeezed states asymptotically tend to a squeezed Bell state 
(corresponding to $\delta_{max}=\frac{\pi}{4}$), which has balanced
Gaussian and non-Gaussian contributions. 

We have compared the non-Gaussianity of the different resources according
to a measure that is reference-dependent. One might think to define the
measure according to an absolute reference. Observing that the squeezed 
Bell-like states and the photon-added/subtracted squeezed states are all
obtained through a degaussification protocol from a pure squeezed state, 
one could modify the definition (\ref{nonGaussianity}) by taking
the twin-beam $|\zeta'\rangle_{12}$ $(\zeta'=r'
e^{i\phi'})$ as the universal reference Gaussian state $\rho_{G}$. 
Adopting this modified definition, and observing that 
the non-Gaussian states to be compared and the reference Gaussian state are 
all pure, Eq. (\ref{nonGaussianity}) reduces to $d_{nG} \,=\, \min_{r',\,\phi'}
\{1-Tr[\rho \, \rho_{G}]\}$, where the minimization is constrained
to run over the squeezing parameters $\zeta'$ of the reference twin-beam.
However, it turns out that this modified definition provides results and 
information qualitatively analogous to those obtained by applying the original 
definition. 

There is still one property that plays a crucial role in the sculpturing
of an optimized CV non-Gaussian entangled resource. From Figs. \ref{vonNeumSBelloptCF}
and \ref{nonGaussMeas} we see that at sufficiently large squeezing the photon-added
and photon-subtracted squeezed resources have entanglement comparable to
that of the optimized squeezed Bell-like states and, moreover, possess stronger
non-Gaussianity. Yet, even in this regime, they are not
able to perform better than the optimized Bell-like resources.
This fact can be understood as follows, leading to the definition
of the {\em squeezed-vacuum affinity}:
It is well known that the Gaussian twin-beam
in the limit of infinite squeezing realizes exactly the CV version of the 
maximally entangled Bell state in the case of qubits. These two ideal resources,
respectively in the CV and qubit case, allow perfect quantum teleportation
with maximal, unit fidelity. Therefore, we argue that, even when exhibiting 
enhanced properties of non-Gaussianity and entanglement, any efficient resource
for CV quantum information tasks should enjoy a further property, i.e. to
resemble the form of a two-mode squeezed vacuum, as much
as possible, in the large $r$ limit.

The {\em squeezed-vacuum affinity} can be quantified by the following maximized overlap:
\begin{equation}
\mathcal{G} \,=\, \max_{s} |\,_{12}\langle -s|\psi_{res}(r)\rangle_{12}|^{2} \; ,
\label{OverlapTWB}
\end{equation}
where $|-s\rangle_{12}$ is a two-mode squeezed vacuum with real squeezing
parameter $-s$, and $|\psi_{res}(r)\rangle_{12}$ is any entangled two-mode 
resource that depends uniquely on the squeezing $r$ as the only free parameter.
This definition applies straightforwardly to the photon-added and photon-subtracted
squeezed resources, and as well to the squeezed Bell-like resources after
optimization with respect to the input state.
The maximization over $s$ is imposed in order to determine, at fixed $r$, the twin-beam
that is most affine to the non-Gaussian resource being considered. 

\begin{figure}[h]
\centering
\includegraphics*[width=7cm]{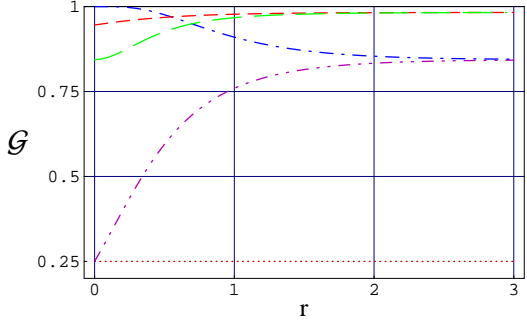}
\caption{(color online) Maximized overlap $\mathcal{G}$ between a twin beam and different non-Gaussian
entangled resources $|\psi_{res}(r)\rangle_{12}$ as a function of $r$. Dashed line:
squeezed Bell-like state with $delta$ fixed at the optimized value $\delta = \delta_{max}^{(C)}$.
Long dashed line: the same with $delta$ fixed at the optimized value $\delta = \delta_{max}^{(F)}$.
For comparison, we plot as well the maximized overlap with the photon-added squeezed
state (\ref{PAS11}) (double-dotted, dashed line); the photon-subtracted squeezed
state (\ref{PSS11}) (dot-dashed line); and the single-photon squeezed number state (\ref{SN11}) 
(dotted line).}
\label{OverlapTWBRes}
\end{figure}

In Fig. \ref{OverlapTWBRes} we study the behavior of the overlap $\mathcal{G}$ as
a function of the squeezing $r$ for different non-Gaussian entangled resources.
From Fig. \ref{OverlapTWBRes} one observes that maximal affinity, and close to unity, 
is always and beautifully achieved, at large values of the squeezing parameter, 
by the optimized squeezed Bell-like resources, while the lowest, constant affinity is 
always exhibited by the squeezed number states. 

In conclusion, optimized squeezed Bell-like resources are the ones that in all 
squeezing regimes are closest to the simultaneous maximization of entanglement, 
non-Gaussianity, and affinity to the two-mode squeezed vacuum. The optimized 
interplay of these three properties explains the ability of squeezed Bell-like 
states to yield better performances, when used as resources for CV quantum 
teleportation, in comparison both to Gaussian resources at finite squeezing and 
to the standard degaussified resources such as the photon-added and the photon-subtracted 
squeezed states. In the next section we will discuss methods and schemes for the 
experimental production of squeezed Bell-like entangled resources.

\section{Methods for the generation of degaussified and squeezed Bell-like resources}
\label{ExpGeneration}

While two-mode (Gaussian) squeezed states are currently produced
in the laboratory, the experimental generation of non-Gaussian
(nonclassical) states in quantum optics is still a hard task, as
it requires the availability of large nonlinearities and/or the
arrangement of proper apparatus for conditional measurements. 
Nevertheless, some truly remarkable realizations of single-mode 
non-Gaussian states have been recently carried out through the use of 
parametric amplification plus postselection 
\cite{ZavattaScience,ExpdeGauss1,ExpdeGauss2}.
Recently, by a generalization of the experimental setup used
in Ref. \cite{ExpdeGauss2} to a two-mode configuration, Kitagawa
{\it et al.} proposed a method for the generation of a certain
class of two-mode photon-subtracted states \cite{KitagawaPhotsub}. 

\begin{figure}[h]
\centering
\includegraphics*[width=8.6cm]{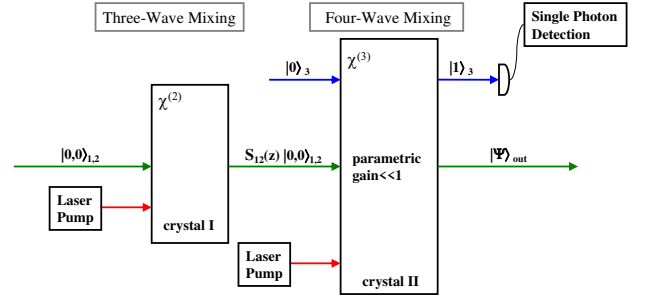}
\caption{(color online) Scheme for the generation of the photon-added squeezed
state (\ref{PAS11}) and of the photon-subtracted squeezed state (\ref{PSS11}).
Two nonlinear crystals are used in a cascaded configuration.
The first $\chi^{(2)}$-crystal is part of a three-wave mixer,
acting as a parametric amplifier for the production
of a two-mode squeezed state $|\zeta\rangle$.
The squeezed state seeds the successive nonlinear process,
a four-wave mixing interaction occurring in a $\chi^{(3)}$-crystal.
A final conditional measurement reduces the multiphoton state
to a photon-added/subtracted squeezed state $|\Psi\rangle_{out}$.}
\label{FigNonGaussGen}
\end{figure}

Here, in some analogy with Ref.
\cite{ZavattaScience}, we propose a possible experimental setup
for the generation of the states (\ref{PAS11}) and (\ref{PSS11}),
and of the squeezed Bell-like states (\ref{squeezsuperpos}).
The scheme, based on a configuration of cascaded crystals, is
depicted in Fig. \ref{FigNonGaussGen}.
In the first stage, by means of a three-wave mixer, functioning as a
parametric amplifier, a two-mode squeezed state 
$|\zeta \rangle = S_{12}(\zeta) |0,0 \rangle_{12}$ is produced. 
In the second stage, a four-wave mixing process takes place in a crystal with
third order nonlinear susceptibility $\chi^{(3)}$. We consider two
possible multiphoton interactions, in the travelling wave configuration, 
described by the following Hamiltonians:
\begin{eqnarray}
H_{I}^{(A)} \,&=&\, \kappa_{A}
a_{1}^{\dag}a_{2}^{\dag}a_{3}^{\dag} \,+\, \kappa_{A}^{*}
a_{1}a_{2}a_{3} \;, \label{Hint4wmA}
\\ && \nonumber \\
H_{I}^{(B)} \,&=&\, \kappa_{B} a_{1}a_{2}a_{3}^{\dag} \,+\,
\kappa_{B}^{*} a_{1}^{\dag}a_{2}^{\dag}a_{3} \;, \label{Hint4wmB}
\end{eqnarray}
where $a_{i}$ $(i=1,2,3)$ denotes three quantized modes of the radiation
field. The complex parameters $\kappa_{A}$ and $\kappa_{B}$ are proportional
to the third order nonlinearity and to the amplitude of an intense
coherent pump field, treated classically in the regime of
parametric approximation. The two-mode squeezed state seeds modes
$1$ and $2$; mode $3$ is initially in the vacuum state
$|0\rangle_{3}$; mode $4$ is the classical pump. Energy conservation 
and phase matching are assumed throughout. Let us remark that, due to 
the typical orders of magnitudes of the third order susceptibilities, the parametric
gains are very small $|\kappa_{A}|\,, \;\;|\kappa_{B}|\ll 1$. The
propagation (time evolution) in the crystal yields
$|\Psi_{I}^{(L)}\rangle \,=\, \exp\{-i t \,
H_{I}^{(L)}\}|\zeta\rangle_{12} \, |0\rangle_{3}$ $(L=A,B)$.
Truncating the series expansion of the evolution operator at the
first order in $\tilde{\kappa}_{L}=-i t \kappa_{L}$, we get
\begin{eqnarray}
|\Psi_{I}^{(A)}\rangle \,&\approx &\, \{1 \,+ \,\tilde{\kappa}_{A}
a_{1}^{\dag}a_{2}^{\dag}a_{3}^{\dag} \} \, |\zeta\rangle_{12} \,
|0\rangle_{3} \label{tevHint4wmA} \,,
\\ && \nonumber \\
|\Psi_{I}^{(B)}\rangle \,&\approx &\, \{1 \,+\, \tilde{\kappa}_{B}
a_{1}a_{2}a_{3}^{\dag} \} \, |\zeta\rangle_{12} \, |0\rangle_{3}
\label{tevHint4wmB} \,.
\end{eqnarray}
Finally, a conditional measurement is performed on mode $3$,
consisting in a single-photon detection, i.e. a projection onto
the state $|1\rangle_{3}$. The postselection reduces the states
(\ref{tevHint4wmA}) and (\ref{tevHint4wmB}) respectively to the states
(\ref{PAS11}) and (\ref{PSS11}). It is worth noting
that the low values of the parametric gains do not affect the
implementation of the process. In fact, it is analogous to
require low reflectivity of a beam splitter to generate 
photon-addition/subtraction by using linear optics. \\
Regarding the production of the squeezed number state (\ref{SN11}), it can be
generated, in principle, by seeding a parametric amplifier with
single-photon states in the two modes. \\
Let us now turn to the experimental generation of the squeezed Bell-like states
(\ref{squeezsuperpos}). They can be engineered by using the same
setup illustrated in Fig. \ref{FigNonGaussGen}, and by simultaneously realizing inside
the nonlinear crystal the processes corresponding to the interactions (\ref{Hint4wmA}) 
and (\ref{Hint4wmB}). In this case the fundamental requirements are that 
of energy conservation and phase-matching for each multiphoton interaction 
must hold simultaneously at each stage.
This conditions can be satisfied by suitably exploiting the phenomenon of birefringence
in a negative uniaxial crystal \cite{MidwinterWarner}.
In particular, the following set of equations must hold:
\begin{eqnarray}
&& \Omega_{1} =  \omega_{1}+\omega_{2}+\omega_{3} \; , \nonumber \\
&& K_{1}^{ext} =  k_{1}^{ord}+k_{2}^{ord}+k_{3}^{ext} \; ,
\label{cond1proc} \\ 
&& \nonumber \\
&& \Omega_{2}+\omega_{1}+\omega_{2} = \omega_{3} \; , \nonumber \\
&& K_{2}^{ord} + k_{1}^{ord}+k_{2}^{ord} = k_{3}^{ext} \; ,
\label{cond2proc}
\end{eqnarray}
where $\omega_{j}$ and $k_{j}^{\lambda}$ $(j=1,2,3)$ represent the frequencies and the
wave vectors of the quantized modes with polarization $\lambda$;
$\Omega_{j}$ and $K_{j}^{\lambda}$ $(j=1,2)$ represent the frequencies and the
wave vectors of the classical pump fields; the superscript $ord$ and $ext$ denote,
respectively, the ordinary and extraordinary polarizations for the propagating waves.
A collinear configuration is assumed for the geometry of propagation inside the crystal.
Then, at fixed $\omega_{1}$ and $\omega_{2}$, the energy conservation relations,
the type-II phase matching condition in Eq. (\ref{cond1proc}),
and the type-I phase matching condition in Eq. (\ref{cond2proc})
can be, in principle, satisfied by a suitable choice of $\omega_{3}$, $\Omega_{1}$, $\Omega_{2}$,
and of the phase-matching angle between the direction of propagation 
and the optical axis. Various examples of such simultaneous multiphoton processes 
have been demonstrated both theoretically and experimentally
\cite{PhysRep,SimultProc1,SimultProc2,SimultProc3,Pfister,Olsen}.
The final conditional measurement on mode $3$ yields the superposition state
\begin{eqnarray}
|\Psi_{I}\rangle & \approx & \tilde{\kappa}_{A} 
a_{1}^{\dag}a_{2}^{\dag} S_{12}(\zeta) |0,0\rangle_{12} \nonumber \\
& + &
\tilde{\kappa}_{B} a_{1}a_{2} S_{12}(\zeta) |0,0\rangle_{12} \; .
\label{psiSimuint}
\end{eqnarray}
By applying a standard Bogoliubov transformation and after a little algebra, 
it is straightforward to show that superposition state (\ref{psiSimuint})
reduces to the squeezed Bell-like state (\ref{squeezsuperpos}) if
\begin{eqnarray}
\label{condizioni}
c_{1} & = & \, -(e^{-i \phi}\tilde{\kappa}_{A}\tanh r +
e^{i \phi}\tilde{\kappa}_{B}) \; , \nonumber \\
&& \nonumber \\
c_{2} & = &
\tilde{\kappa}_{A} + e^{2i \phi}\tilde{\kappa}_{B} \tanh r \; .
\end{eqnarray}
The latter conditions can be successfully implemented by observing
that the complex parameters $\tilde{\kappa}_{A}$ and $\tilde{\kappa}_{B}$ 
can be controlled to a very high degree by means of the amplitudes
of the external classical pumps.

\section{Conclusions and outlook}
In this work we have presented a thorough comparison, with regard to
the performance in continuous-variable quantum teleportation, between 
standard degaussified resources such as photon-added and photon-subtracted 
squeezed states and a new type of {\em sculptured} resource that interpolates
between different degaussified states and can be optimized because it depends 
on an extra, relative-phase, independent free parameter in addition to squeezing.  
These sculptured non-Gaussian resources are what we have named {\em squeezed Bell-like 
states}: They hybridize discrete single-photon pumping, coherent superposition of 
Bell two-qubit eigenstates, and continuous-variable squeezing.
The maximization of the teleportation fidelity with respect to different inputs,
including coherent and squeezed states, is achieved by squeezed Bell-like states
in comparison both to Gaussian and other non-Gaussian resources, and for all
values of squeezing, including the asymptotic Einstein-Podolsky-Rosen limit.
Understanding this enhancement yielded by the squeezed Bell-like resources 
in the teleportation success is possible when interpreted in terms of a 
multiple optimization problem. The squeezed Bell-like states are those states
that are as close as possible to the simultaneous maximization of entanglement,
non-Gaussianity, and affinity to the two-mode squeezed vacuum. The analysis
performed in the case of pure-state resources can be extended to the case
of mixed-state resources in the presence of noise, imperfections, and other
sources of decoherence: We plan to report the results on the study of these
situations in the near future.

The concepts of hybridization, sculpturing, and optimization suggest that
the present investigation could be extended and generalized along several
directions. Further optimization is in principle possible with respect to 
the local parts of the resource states, in analogy to the case of standard 
Gaussian resources \cite{Telepoppy}. One could think of extending the sculpturing 
to the entire basis of Bell states, to generate entangled non-Gaussian resources 
that can never be reduced to proper truncations of Gaussian squeezed resources.
Such ``fully sculptured'' resources might allow for the further enhancement
of the teleportation success due to the presence of a larger number of
experimentally adjustable free parameters in addition to squeezing.
Fully sculptured states could be applied to hybrid schemes
of teleportation combining continuous-variable inputs with discrete-variable
resources and viceversa. In this framework, a particularly appealing line
of research would be to look for modified schemes of teleportation beyond
the standard Braunstein-Kimble protocol, to be realized by generalized
measurements in combination with state-control enhancing unitary operations. 
Finally, the present discussion could be extended to other types of 
quantum information tasks and processes besides teleportation. For instance,
it would be interesting to investigate the comparative effects of non-Gaussian 
inputs and non-Gaussian resources in schemes for the generation of macroscopic 
and mesoscopic optomechanical entanglement \cite{Aspelmeyer}.

\acknowledgments

We acknowledge financial support from MIUR, under PRIN National Research
Program 2005, from CNR-INFM Coherentia, from CNISM, from INFN, under the special
HALODYST program, and from ISI Foundation for Scientific Interchange.

\end{document}